\documentclass[showpacs,preprintnumbers,amsmath,amssymb]{revtex4}
\usepackage{graphicx}
\usepackage{latexsym}
\usepackage{caption}
\usepackage{capt-of}
\usepackage{amsmath}
\usepackage{exscale,relsize}
\usepackage[toc,page]{appendix}

\begin{document}

\title{Angular Dependence of Coherent Pion Production by Neutrinos}

\newcommand{\Guanajuato}{DCeI Le\'on, Universidad de Guanajuato, C.P. 37150, Le\'on, Guanajuato, M\'exico.
}
\affiliation{\Guanajuato}
\newcommand{\Dortmund}{Department of Physics, Technical University of Dortmund, D-44221, Dortmund, Germany}
\affiliation{\Dortmund}
\author{A.~Higuera}
\email{higuera@fnal.gov}
\affiliation{\Guanajuato}
\author{E. A.~Paschos}
 \email{paschos@physik.uni-dortmund.de}
\affiliation{\Dortmund}

\date{\today}

\begin{abstract}
Coherent pion production by neutrinos has been interpreted in the framework of the Partially Conserved Axial Current hypothesis (PCAC) and explicit model calculations are available. In this article we compute angular correlations for the produced pions which may help to separate the signal from the background. We present many figures useful for the experiments and compare them with another model.
\end{abstract}
\maketitle

\section{Introduction}

In the early 80's the experimental results of neutrino interaction with the matter attracted a lot of interest in attempts to determine the structure of neutral current interactions. For instance the production of a single $\pi^{0}$ in the reaction $\nu +n \rightarrow \nu + n + \pi^{0}$, became very interesting since detectors sometimes register only one of the decay photons, and this constituted a source of background to neutrino-electron scattering and for  $\nu_{\mu} \rightarrow \nu_{e} $ oscillation searches. By 1978 the Aachen-Padova collaboration \cite{Measurement of Muon-Neutrino and Antineutrino Scattering off Electrons Aachen-Padova} measured the $\nu_{\mu}  e $ scattering and found an excess of showers. A few years later the reaction $\nu + N \rightarrow \nu +N +\gamma $, the coherent emission of a photon in a neutrino-nucleus collision, was proposed in order to examine the extent to which the  "excess" of showers seen in the Aachen-Padova experiment could be understood by this mechanism and  by using the PCAC (partially conserved axial current) hypothesis \cite{Coherent Production of Photons by neutrinos}. Studying the results of the Aachen-Padova experiment it was concluded that coherent photons could not account for the excess of events in the very forward direction ($\theta \sim 2^{\circ}$) \cite{Measurement of Muon-Neutrino and Antineutrino Scattering off Electrons Aachen-Padova}. Inspired by the idea of coherent emission of a photon and the PCAC hypothesis, the coherent single $\pi^{0}$ production  in neutrino reactions was introduced  $\nu+A \rightarrow \nu +A +\pi^{0}$ \cite{RS}. The observation of coherent $\pi^{0}$ production in neutrinos was first reported in 1983 by the Aachen-Padova experiment \cite{Coherent Aachen-Padova}.  Subsequent observations of the $\pi^{0}$ production were reported by the Gargamelle collaboration \cite{Gargamelle} 
followed by the CHARM collaboration \cite{Charm1985}. The BEBC WA59 Collaboration reported the $\pi^{-}$ production by antineutirnos \cite{BEBCWA59}, during the same month the SKAT collaboration observed for the frist time in one experiment all three states of the isospin triplet of the axial part of the weak charged and neutral currents \cite{SKAT}. Additional observations  of coherent $\pi^{0}$  production where found in \cite{Fermilab15ftbubble} and \cite{NOMAD} and the latest measurement was reported by ScibooNE collaboration \cite{ScibooneNC}. Coherent $\pi^{\pm}$ pion production where observed in  \cite{BEBCWA59_second} and \cite{CharmII}. However two recent measurements on the charged current case  \cite{K2K} and \cite{Sciboone} with low energy neutrino beams, have not found evidence. The question arises if the effect is absent for the charged current reactions or it is suppressed by the kinematic, especially effects from the muon mass.

In section \ref{sec:Coh} we present three PCAC models pointing out advantages and shortcomings of the implementations for three PCAC methods. In section \ref{sec:angular} we use the formalism of references  \cite{Kartavtsev:2006} and \cite{PS}  in order to calculate angular distributions of the pions relative to the neutrino direction.  We show the ability of  models based on PCAC  to determine  angular distributions of the produced pion relative to the neutrino direction for various value of $Q^{2}.$ The dependence in the polar angle $\theta$ and the azimuthal $\phi$ relative to the direction of the neutrino, peaks at small values of these variables. Finally in section \ref{sec:summary} we present our conclusions and give the cross section $d\sigma/dcos\theta$ comparing two PCAC models and averaging over the neutrino spectrum of the MINERvA experiment.

\section{Coherent Pion Production by Neutrinos in the PCAC Framework}
\label{sec:Coh}

The process of coherent pion production by neutrinos implies that the nucleus does not break up or alter its quantum numbers during the process.
\begin{equation}\label{CCCoh}
\nu_{l} (k) + A(p) \rightarrow l^{-}(k')  + A(p')\ + \pi^{+}(p_{\pi}).
\end{equation}
It follows from this that the momentum transfer to the nucleus cannot be too large, otherwise a struck nucleon will be knocked out.   Because of this the coherent interactions are essentially diffractive, characterized by an exponentially falling momentum transfer to the nucleus distribution and a forward outgoing pion.  The coherent cross section also depends on the atomic mass number of the nucleus.

As we mentioned in Section I The Aachen-Padova collaboration observed an excess of events localized in very forward directions of the neutrino interactions, so that coherent pion production by neutrinos becomes a natural candidate for explaining such interactions and at the same time providing a test for PCAC hypothesis.  In this section we present some of the coherent models based on PCAC and their implications.  

Another class of models introduces a microscopic picture for the production of the pions through diagrams in the delta resonances region, typically  $\Delta$(1232) \cite{Microscopic1, Microscopic2, Microscopic3, Microscopic4, Microscopic5,Microscopic6, Leitner} and also includes background terms. They account for nuclear medium effects with a multiplicative factor, the Fourier transform of the local density for protons and neutrons. There are various versions of these models with local approximation \cite{Microscopic1, Microscopic2, Microscopic3, Microscopic4, Microscopic5,Microscopic6} and without local approximation \cite{Leitner}.  In either case the models are valid at low energies $ E_{\nu}\sim$ 2.0 GeV. A review of the Rein-Sehgal  \cite{RS}, its use at low energies and a comparison with a microscopical calculation can be found in \cite{Microscopic6}.

The Rein-Sehgal model \cite{RS} is often used as input model for Monte Carlo event generators like GENIE \cite{GENIE}. The authors estimate the coherent cross section for neutral current using Adler's PCAC theorem \cite{Adler:1964} and extended in the non-forward directions $(Q^{2} \neq0)$ by adding a propagator term with an axial mass $m_{A}$. The Rein-Sehgal triple differential cross section for the reaction (\ref{CCCoh}) is given by 
\begin{align}\label{RS_eq4}
\frac{d\sigma}{dx\ dy\ dt}= \frac{G^{2}_{F}Mf^{2}_{\pi}A^{2}}{2\pi^{2}}E(1-y)\frac{1}{16\pi}(\sigma^{\pi N}_{tot})^{2} (1+r^{2})  \left(\frac{m^{2}_{A}}{m^{2}_{A}+Q^{2}}\right)^{2}e^{-b|t|}F_{abs},
\end{align}
where $x = Q^{2}/2M\nu$, $y= \nu/E$,  $t$ is the momentum transfer-squared to the nucleus defined as $t=(p_{\pi}-q)^{2}$, $\sigma^{\pi N} $ is the pion-nucleon cross section, $M$ denotes the nucleon mass, $A$ is the number of nucleons within the nucleus, $r$ defined as the ratio of the real to imaginary part of the forward pion-nucleon scattering amplitude $r = Re f_{\pi N}(0)/Im f_{\pi N}(0)$, $b = 1/3R^{2}$, $(R= R_{0}A^{1/3})$ with $R$ the nuclear radius and $F_{abs}$ given by
\begin{equation}\label{RS_eq5}
F_{abs} = exp\left\{-\frac{9A^{1/3}} {16\pi R^{2}_{0}}\sigma^{\pi N}_{inel} \right\}.
\end{equation}
Later on the authors in \cite{RS2} made an attempt to reconcile their model with the data from \cite{K2K} by taking into account the mass of the muon as a simple multiplicative correction factor
\begin{equation}\label{RS2_eq1}
C = \left(1-\frac{1}{2}\frac{Q^{2}_{min}}{Q^{2}+m^{2}_{\pi}}\right)^{2}+\frac{1}{4} y \frac{Q^{2}_{min}(Q^{2}-Q^{2}_{min})}{(Q^{2}+m^{2}_{\pi})^{2}},
\end{equation}
where
\begin{equation}\label{RS2_eq2}
Q^{2}_{min} = m^{2}_{l}\frac{y}{1-y},
\end{equation}
with the range $Q^{2}_{min} \leq Q^{2} \leq 2mEy_{max}$  where $y$ lies between $y_{min} = m_{\pi}/E$ and $y_{max} = 1 - m_{l}/E$. Thus the new result is given by
\begin{align}\label{RS2_eq3}
\frac{d\sigma}{dxdydt}= 2\left(\frac{d\sigma^{\pi^{0}}}{dxdydt}\right)C\theta(Q^{2}-Q^{2}_{min}) \theta(y-y_{min})\theta(y_{max}-y),
\end{align}
which according to the authors in \cite{RS2} gives a 25$\%$ suppression of the cross section caused by a destructive interference of the axial vector and pseudoscalar amplitudes.

There is an updated version of the original Rein-Sehgal model and its extension to charged current reactions made by Berger and Sehgal \cite{Berger:2009}, where instead of using models for pion nucleus scattering as used in \cite{RS}, the available data on pion-carbon scattering are implemented in the numerical analysis. The Berger-Sehgal triple differential cross section 
for reaction (\ref{CCCoh}) is given by
\begin{align}\label{BS_eq}
\frac{d \sigma}{dQ^{2}\ dy \ dt}= \frac{G^{2}_{F}cos^{2}\theta_{c}f^{2}_{\pi}}{2\pi^{2}}\frac{E}{| \vec{q} |}uw \bigg{[}\left(G_{A}-\frac{1}{2}\frac{Q^{2}_{min}}{Q^{2}+m^{2}_{\pi}}\right)^{2} 
+   \frac{y}{4}(Q^{2}-Q^{2}_{min})\frac{Q^{2}_{min}}{Q^{2}+m^{2}_{\pi}} \bigg{ ]}  \frac{d\sigma^{\pi A}}{dt},
\end{align}
with $u, w = (E+E' \pm | \vec{q} |)/2E$, $\theta_{c}$ the Cabbibo angle, $Q^{2}_{min} = m^{2}_{l}y/(1-y)$, $G_{A} = m^{2}_{A}/(Q^{2}+m^{2}_{A})$ and $d\sigma^{\pi A}/dt$ is the differential pion nucleus cross section. Another contribution of the authors in \cite{Berger:2009} is to replace the original phenomenological expression of the differential pion nucleus scattering with the ansatz

\begin{equation}\label{BS_anatz}
\frac{d\sigma^{\pi A}}{dt} = A_{1}e^{-b_{1}|t|},
\end{equation}
with energy dependent coefficients $A_{1}$, $b_{1}$ which are listed in Table I of reference \cite{Berger:2009}. Even though these authors claim to include muon mass effects the product $uw$ in Eq. (\ref{BS_eq}) neglects the mass of the muon. 

Another model especially suited for charged current coherent scattering  has been developed by Kartavtsev, Paschos and Gounaris \cite{Kartavtsev:2006} and  extended by Paschos and Schalla  \cite{PS}. As in the previous model $\pi A$ scattering data are used. This model incorporates the pion-nucleus cross section into the neutrino scattering and uses the lepton mass exactly. 
\begin{align}\label{PS_eq5}
\frac{d\sigma}{dQ^{2}d\nu dt}=  \frac{G^{2}_{F}|V_{ud}|^{2}}{2(2\pi)^{2}} \frac{\nu f^{2}_{\pi}}{E^{2}Q^{2}} \bigg{\{} \tilde{L}_{00}+\tilde{L}_{ll}\left( \frac{m^{2}_{\pi} } {Q^{2} +m^{2}_{\pi}} \right)^{2}
+ 2\tilde{L}_{l0}\frac{ m^{2}_{\pi} }{Q^{2}+m^{2}_{\pi}} \bigg{\}} \frac{d\sigma^{\pi A}}{dt}.
\end{align}
With $|V_{ud}|^{2}$ the magnitude of  the CKM matrix elements, $d\sigma^{\pi A}/dt $ being the elastic pion-nucleus cross section and $\tilde{L}_{ij} = 1/2 L_{ij}$.  The authors parametrized the elastic pion nucleus cross section also with Eq. (\ref{BS_anatz}), fitting the parameters $A_{2}(E_{\pi})$ and $b_{2}(E_{\pi})$ to the $\pi^{+}C^{12}$ elastic scattering data. The density matrix elements are given by

\begin{equation}\label{PS_eq2}
L_{00} = 4 \left\{ \frac{[Q^{2}(2E-\nu)-\nu m^{2}_{l} ]^{2}}{Q^{2}(Q^{2}+\nu^{2})} -(Q^{2}+m^{2}_{l})\right\},
\end{equation}
\begin{equation}\label{PS_eq3}
L_{l0} = 4m^{2}_{l}\frac{Q^{2}(2E-\nu)-\nu m^{2}_{l}}{Q^{2}\sqrt{Q^{2}+\nu^{2}}},
\end{equation}
\begin{equation}\label{PS_eq4}
{L}_{ll}= 4m^{2}_{l}\left(1+\frac{m^{2}_{l}}{Q^{2}}\right).
\end{equation}
As we mentioned above the $L_{00}$ corresponds to the $uw$ term in Eq. (\ref{BS_eq}) and is proportional to it when we neglect the lepton mass. Taking the divergence of the spin averaged matrix element and invoking PCAC hypothesis, the authors write the matrix element of the axial  current as the sum of the pion pole and remaining contribution and estimated the matrix element. The corresponding neutral current cross section is
\begin{equation}\label{PS_NC}
\frac{d\sigma}{dQ^{2}d\nu dt}=  \frac{G^{2}_{F}|V_{ud}|^{2}}{4(2\pi)^{2}} \frac{\nu f^{2}_{\pi}}{E^{2}Q^{2}}\left\{ \tilde{L}_{00}\right\}\frac{d\sigma^{\pi A}}{dt}.
\end{equation}
In this formula the muon mass in $ \tilde{L}_{00}$ must to be set zero.

The three approaches start from the same physical motivation and rely on the principe of applying PCAC. They arrive at results which differ because they implement different methods. The differences occur at low neutrino energies and disappear for energies higher than 3.0 GeV. It is important to find out for charged current reactions whether coherent scattering occurs and at which level.  We calculate the angular distribution of the pion relative to the neutrino direction, which can be used to separate the signal from the incoherent background.

\section{Angular Dependence}
\label{sec:angular}

\begin{figure*}[t!]
\begin{center}
\includegraphics[width=0.8\textwidth,height=3.0in]{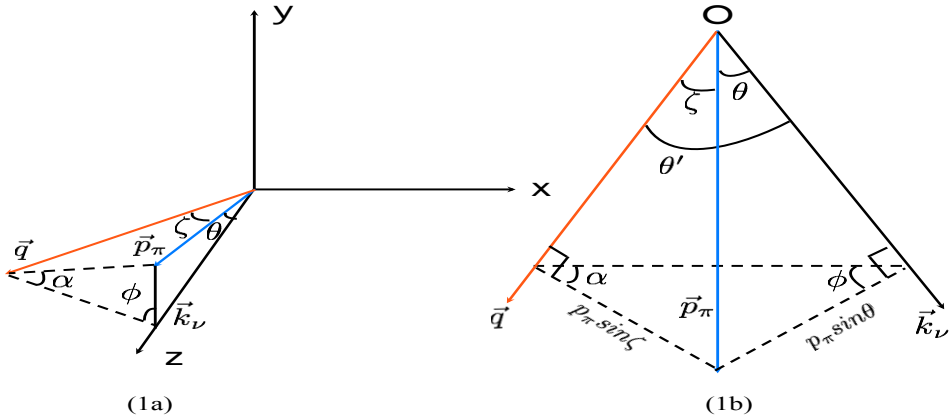}
\caption{Coordinate system used for the angular dependence. The vector $\vec{p}_{\pi}$ is, in general outside of the x-z plane; the coordinates of $\vec{p}_{\pi}$ relative to $\vec{q}$ are defined by $\zeta$ and $\alpha$. Whereas the coordinates of the $\vec{p}_{\pi}$ relative to the neutrino direction are defined by $\theta$ and $\phi$. The tetrahedron in (1b) is seen from above with $\vec{k}_{\nu}$ and $\vec{q}$ lying on the page and forming an angle $\theta'$ between them.} 
\label{fig:diagram}
\end{center}
\end{figure*}

The angular dependence of the process is very characteristic and it should be helpful for separating the signal from the background. We define a right-handed coordinate system with the neutrino direction along the z-axis  Fig. \ref{fig:diagram}a. The weak current lies on the x-z plane and makes an angle $\theta'$ relative to the direction of the neutrino. Thus the x-z plane will be called the lepton plane, since the muon also lies on this plane. The coordinates of the $\vec{p}_{\pi}$ relative to the neutrino direction is defined by the polar angle $\theta$ and the azimuthal $\phi$; and relative to $\vec{q}$ are defined by $\zeta$ and $\alpha$ respectively. The vector $\vec{p}_{\pi}$ and its projections to $\vec{k}_{\nu}$ and $\vec{q}$ define a tetrahedron, shown in Fig. \ref{fig:diagram}b with two right angles. In this coordinate system
\begin{equation}
\vec{q}=|\vec{q}|( cos\theta' \hat{k}-sin\theta' \hat{i})
\end{equation}
and 
\begin{equation}
\vec{p}_{\pi} = |\vec{p}_{\pi}| (cos\theta \hat{k}+sin\theta(-cos\phi \hat{i}+sin\phi \hat{j} )).
\end{equation}
The momentum transfer squared between the neutrino and the muon $q^{2} = (k'-k)^{2}=-Q^{2}$ and in the limit $\nu^{2} \gg Q^{2}$ is approximated by 
\begin{equation}\label{Q2_eq}
Q^{2} \approx \frac{2EE^{2}_{\pi}}{Ecos\theta'-E_{\pi}}(1-cos\theta').
\end{equation}

The geometry of the tetrahedron Fig. \ref{fig:diagram}b gives useful relations. For instance, the angle $\alpha$ is related to the others by 
\begin{equation}\label{alpha}
sin\alpha = \frac{sin\phi}{sin\zeta}sin\theta,
\end{equation}
and the three vectors $\vec{k}_{\nu}, \ \vec{q}$ and $\vec{p}_{\pi}$ satisfy the addition theorem 
\begin{equation}
cos\zeta = cos\theta cos\theta' + sin\theta sin\theta' cos\phi.
\end{equation}
Similarly, the momentum transfer to the nucleus is given by
\begin{equation}
t = (p_{\pi}-q)^{2} = -Q^{2} - 2\nu E_{\pi}+2|\vec{q}||\vec{p}_{\pi}|cos\zeta + m^{2}_{\pi}.
\end{equation}
The original cross section in Eq. (\ref{PS_eq5}) is isotropic on the rotation around the vector $\vec{q}$. To convert the variables of the pion around the neutrino direction, we calculate the Jacobian 
\begin{equation}\label{Jacobian}
dQ^{2}d\nu dt d\alpha = J(\theta, \theta' , \phi)dE_{\pi}d\theta d\theta' d\phi,
\end{equation}
with $J(\theta, \theta' , \phi)$ given in the Appendix. The two cross section are related
\begin{equation}
\frac{d\sigma}{d\nu dQ^{2} dt \frac{d\alpha}{2\pi}} = \frac{d\sigma}{dE_{\pi}J(\theta,\theta',\phi) d\theta d\theta' \frac{d\phi}{2\pi}}.
\end{equation}
They lead to the result 
\begin{align}
\label{angular_3}
\frac{d\sigma}{dE_{\pi}d\theta d\phi} =\frac{G^{2}_{F}|V_{ud}|^{2}}{2(2\pi)^{3}}  A_{2}(E_{\pi}) \int   \frac{\nu f^{2}_{\pi}}{E^{2}Q^{2}}\bigg \{   \tilde{L}_{00}   + \tilde{L}_{ll} \left( \frac{m^{2}_{\pi} } {Q^{2} +m^{2}_{\pi}} \right)^{2}   +\  2\tilde{L}_{l0}\frac{ m^{2}_{\pi} }{Q^{2}+m^{2}_{\pi}} \bigg \} e^{-b_{2}|t|}J(\theta, \theta', \phi)d\theta'.
\end{align}

Among the terms in the curly brackets  the largest contribution comes from the term $L_{00}$ (helicity zero contribution). This is clear by looking at the numerical values of the three density-matrix elements which we calculated and present in Table \ref{table} in the Appendix. 

As a first presentation of numerical results we consider the case when the pion lies on the leptonic plane, {\em i.e.} $\phi=0$, For this case the Jacobian is given by
\begin{equation}\label{Jacobian_phi0}
J(\theta,\theta', \phi =0) = \frac{4EE^{2}_{\pi}}{E-E_{\pi} } |\vec{q}| |\vec{p}_{\pi}| sin\theta' sin\theta.
\end{equation}

\begin{figure*}[h!]
\begin{center}
\includegraphics[width=1\textwidth,height=7in]{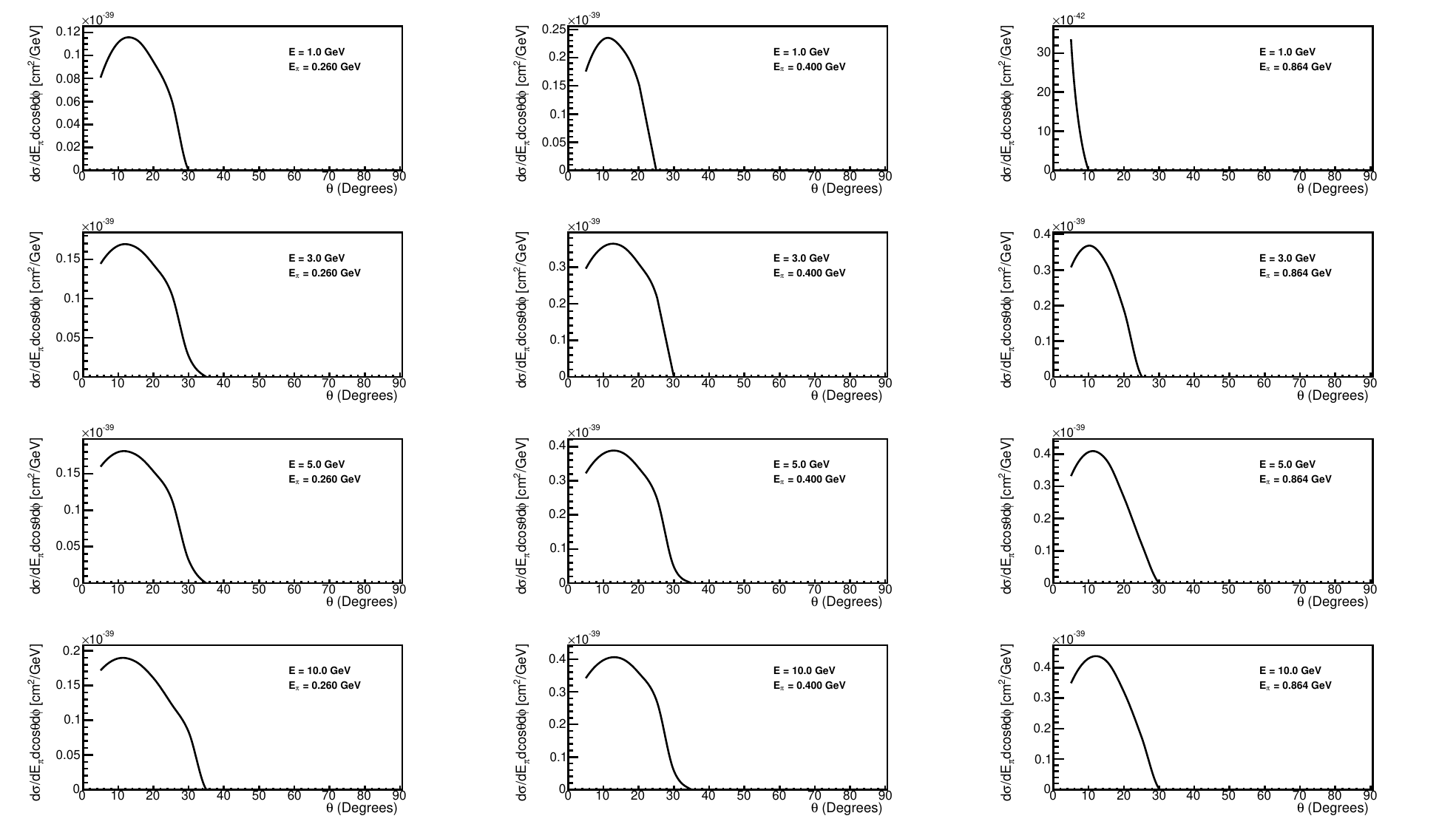}
\caption{\small Triple differential cross section Eq. \ref{new_XS}, case $\phi$ = 0. }
\label{fig:triple_phi0}
\end{center}
\end{figure*}

The simple geometry gives the relation $\theta' = \zeta+\theta$ which permits a simple change of variables from $\theta'$ to $\zeta$. For the pion-Carbon cross section we use the functional form Eq.  (\ref{BS_anatz}) and the numerical values for $A_{2}(E_{\pi})$ and $b_{2}(E_{\pi})$ are taken from the Table I in Ref. \cite{PS}. For values of $E_{\pi}>$ 1.046 GeV the pion-Carbon cross section was extrapolated as constant {\em i.e.}  using the last numerical value given by \cite{PS}. Keeping the leading term $L_{00}$ we have
\begin{align}\label{new_XS}
\frac{d\sigma}{dE_{\pi}dcos\theta d\phi} = \frac{G^{2}_{F}|V_{ud}|^{2}}{2(2\pi)^{3}}  \frac{4EE^{2}_{\pi}}{E-E_{\pi} }A_{2}(E_{\pi}) |\vec{p}_{\pi}|  \int  |\vec{q}| \left(\frac{\nu \tilde{L}_{00} f^{2}_{\pi}}{E^{2}Q^{2}}\right) e^{-b_{2}|t|}sin(\zeta+\theta) d\zeta.
\end{align}
We carried out the integration numerically imposing at every step the $|t_{min}|$  condition $|t_{min}|\geq \left( \frac{Q^{2}+m^{2}_{\pi}}{2\nu}\right)^{2}$, that was introduced by the authors in \cite{Kartavtsev:2006}. This condition is important and must be used in all models.

\begin{figure*}[!t]
\begin{center}
\includegraphics[width=1\textwidth,height =7in]{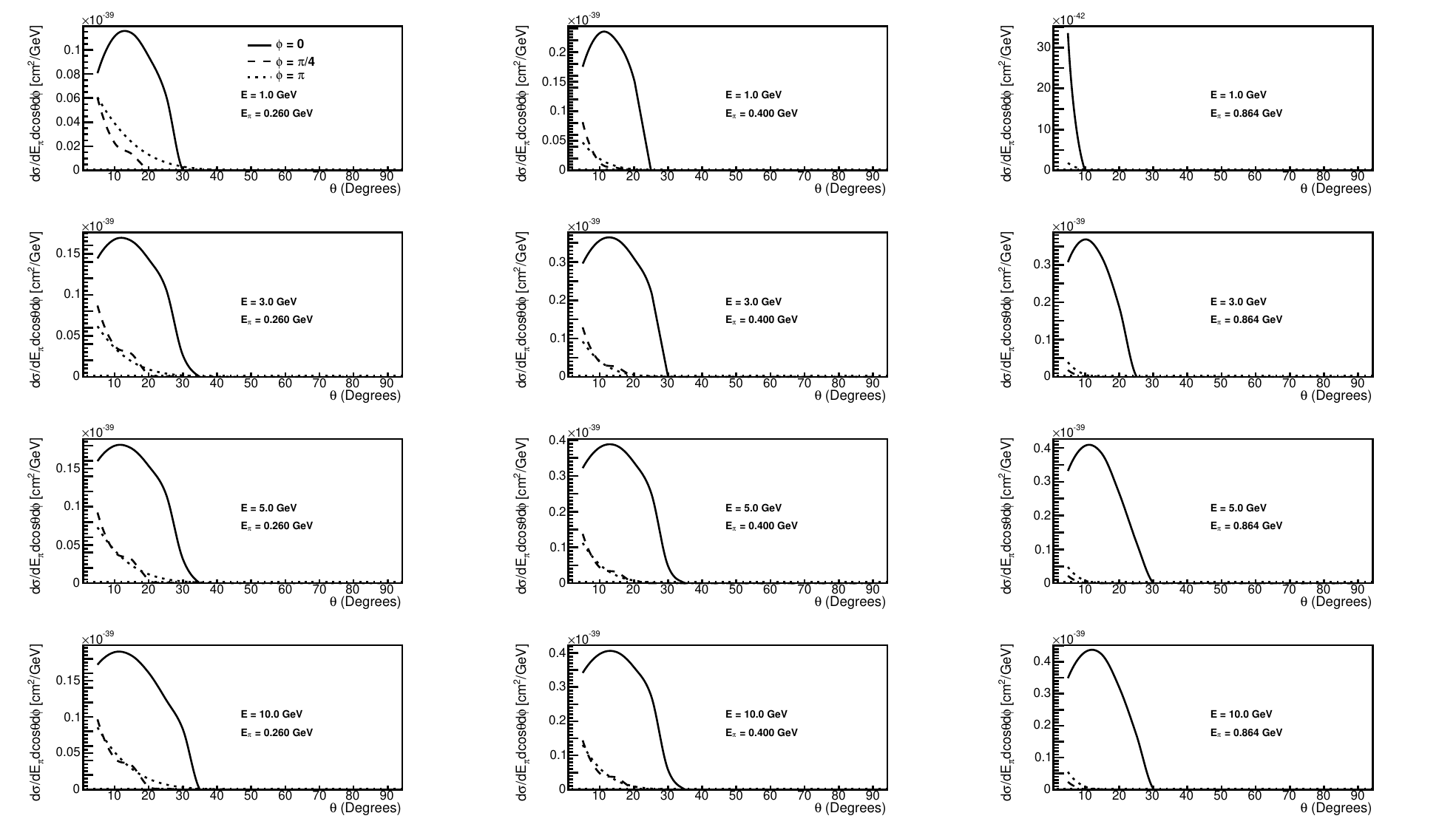}
\caption{\small Triple differential cross section Eq. \ref{new_XS}, for $\phi$ = 0, $\pi/4$ and $\pi$, the cross section for  $\phi=\pi/2$  is negligible compared to the others.  }
\label{fig:triple_phi}
\end{center}
\end{figure*}

We consider the following physical situation; we keep the energy and angles of the pion, {\em i.e.} $\theta$ and $\phi$, fixed and sum over the directions of the current. This is equivalent to the integration over the angle $\theta'$ with the appropriate limits. We obtained angular distributions for neutrino energies  E=1, 3, 5, 10 GeV and values of $E_{\pi}$ = 0.260, 0.400, 0.864  GeV for $\phi$ = 0. We present the results in Fig. \ref{fig:triple_phi0}. We notice that the cross section is concentrated at small values of $\theta$; in fact for $\theta > 30^{\circ}$ there are hardly any events. The sharp dependence on $\theta$ reflects the sharp peak of the pion-Carbon cross section. This is also the characteristic peaking observed in the early experiments reviewed in the introduction. For fixed $E$ the distribution in $\theta$ becomes sharper as $E_{\pi}$ increases.

We want to point out that the sharp decrease of $e^{-b_{2}|t|}$ within the integrand depends on the value of $b_{2}(E_{\pi})$ and influences the normalization and not so much the shape of the curves. The cross section we present in figures \ref{fig:triple_phi0}, \ref{fig:triple_phi} have absolute normalizations which change when we change the input pion-Carbon data.

The $\phi$ dependence is also very peaked at small values of $\phi$. In Fig. \ref{fig:triple_phi} we show curves for three values of $\phi =0, \pi/4$ and $\pi$ it is evident that most of the pions are produced in the lepton plane in the opposite side from the muon relative to the neutrino. This is what was expected because the pions in this configuration are closer to the vector $\vec{q}$ and making the $|t|$ variable small.

It is interesting to compare our results with experimental data, however the lack of data makes it difficult. In Ref. \cite{SKAT} Fig. 8a gives an azimuthal distribution, for $|t|\leq 0.15$ GeV$^{2}$ but their variable $\phi$ is different from our definition because they define it relative to the direction of $\vec{q}$. In our definition $\phi$ is relative to the neutrino direction. Besides the different definitions a quantitative comparison is still possible because their case for $\phi = 0$ corresponds to the configuration where the pion is between the neutrino and the current, which is also our configuration for $\phi$= 0  (see Fig. \ref{fig:diagram}). Independently of the differences both cases peak at $\phi=0$ which is encouraging. Another experiment, SciBooNE \cite{Hiraide}, presented a $\Delta \phi$ distribution for $\theta < 35^{\circ}$ where the data shows a peak around $\Delta\phi \sim 0.0^{\circ}$.
  
The main background for coherent scattering consists of events where the nucleus breaks up with a pion and a proton or neutron knocked out but still remaining undetected. The scattering in this case takes place on single nucleons and the kinematics for the reaction are different. In addition both vector and axial currents contribute as well as their interference term. The $t$ dependence of the new process is again exponential but with a much smaller value for $b_{2}(E_{\pi})$. The net effect is a $\theta$ dependence that extents to larger values of the angle.

\begin{figure*}[t!]
\begin{center}
\includegraphics[width=0.48\textwidth,height =3in]{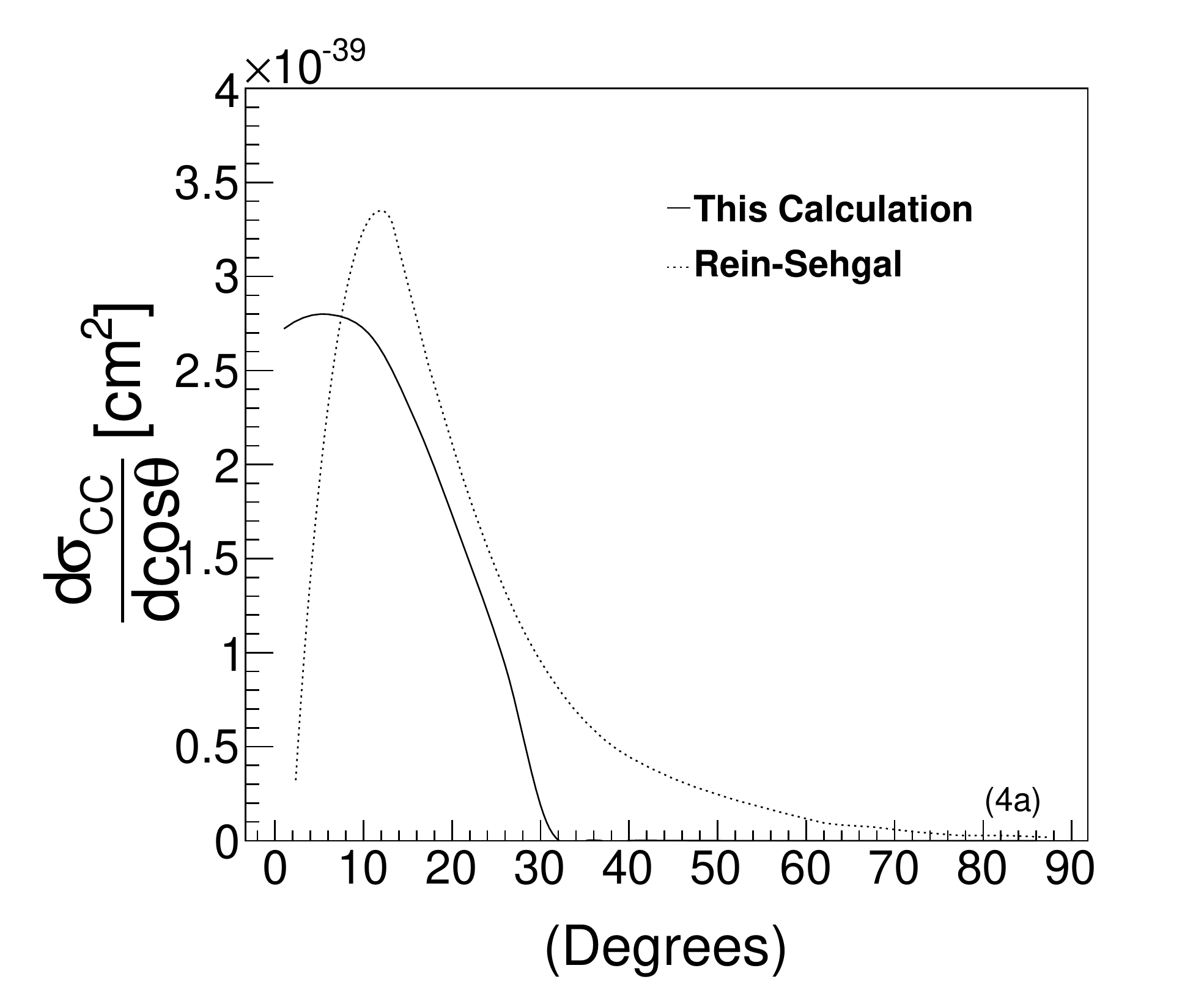}
\includegraphics[width=0.48\textwidth,height =3in]{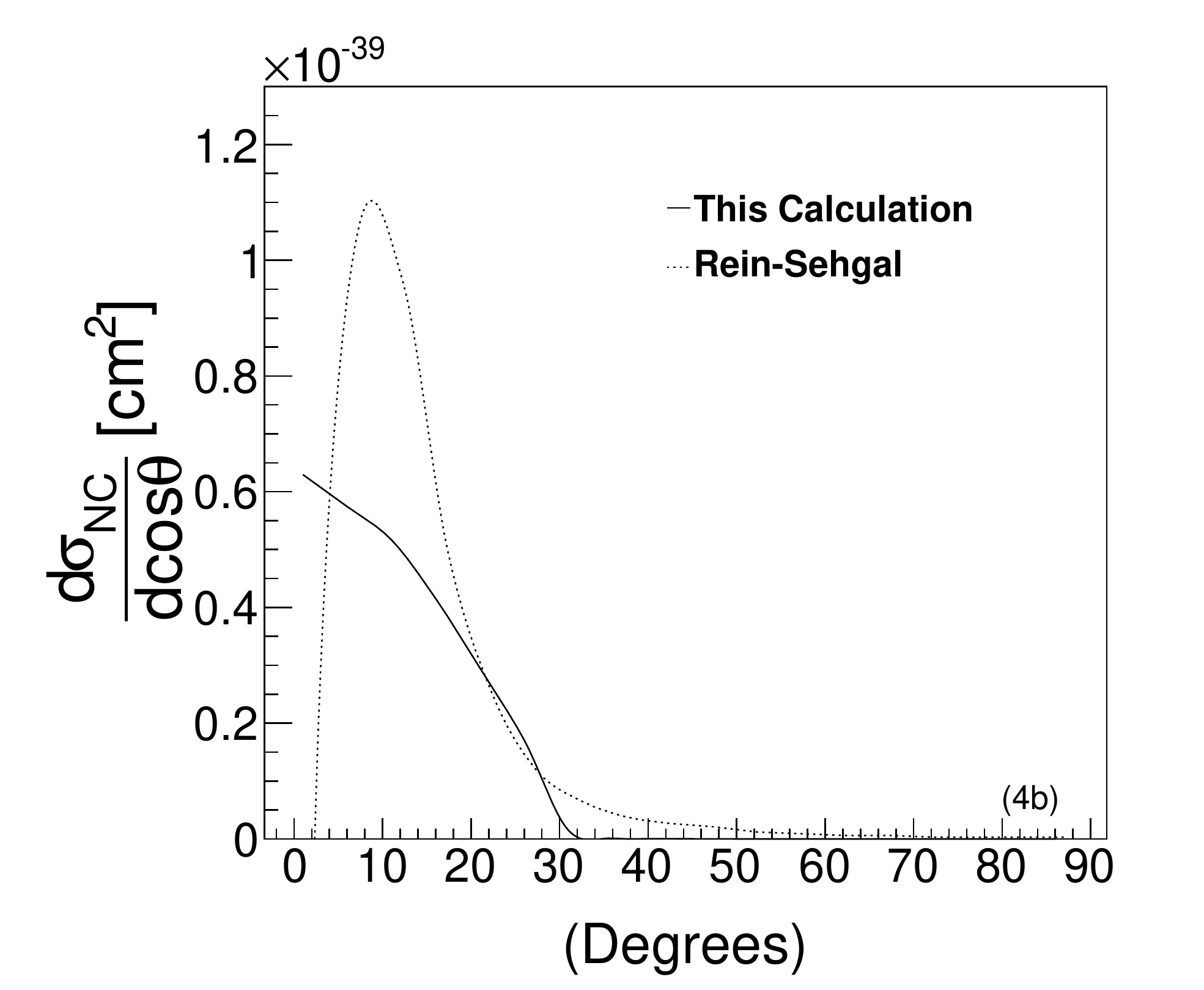}
\caption{\small The differential cross section $d\sigma/dcos\theta$ as a function of $\theta$. Figure (4a) shows the charged current reaction $\nu_{\mu}+C^{12}\rightarrow \mu^- + C^{12}+ \pi^+$ and (4b) is the neutral current reaction $\nu_{\mu}+C^{12}\rightarrow \nu_\mu+C^{12}+\pi^0$. The solid curves are for the calculation described in this article and the dotted curves were obtained using \cite{RS,RS2}. }
\label{fig:ds_dcos}
\end{center}
\end{figure*}

\section{Summary}
\label{sec:summary}

Coherent pion production by neutrinos was studied for CC and NC reactions with special attention to conditions that must be satisfied at lower energies. In this work we summarized formulas and then extended them by calculating angular distributions in the polar angle $\theta$ and azimuthal angle $\phi$.  The physical configurations we considered correspond to the selection of a specific energy $E_{\pi}$ and specific direction of the pion relative to the neutrino direction (specified by fixed values of $\theta$ and $\phi$); then we integrated over the direction of the momentum $\vec{q}$ (momentum of the weak current) keeping the energy  $q_{0} = \nu$ equal to the $E_{\pi}$ as required for coherent scattering. We found a very sharp peak in the variable $\theta$ and also peaking in $\phi$ where small values of $\phi$ are preferred.

In addition we computed the differential cross section  $d\sigma/dcos\theta$ by integrating over $E_{\pi}$ up to 3.0 GeV using again data from Ref. \cite{PS} and for values of $E_{\pi}>1.0$ GeV the pion-Carbon cross section was extrapolated using the last numerical values in Table 1 of \cite{PS}. The results for the charged and neutral currents are shown with solid curves in Fig. \ref{fig:ds_dcos} averaged over the neutrino flux for the MINERvA experiment \cite{Fiorentini}. In order to check the interpolation for $E_{\pi}>1.0$ GeV we reduced the values for $b_{2}(E_{\pi})$ from 53.5 (1/GeV$^{2})$ to 40.0 and 30.0 (1/GeV$^{2}$) and found that the values for the differential cross section hardly change. The high energy range of integration where $E_{\pi}> 1.0$ GeV, the cross section is smaller and with a sharper distribution in $\theta$ (see Fig. \ref{fig:triple_phi}) for this reason the change in $d\sigma/dcos\theta$  is very small. We also computed the same cross section for the model \cite{RS,RS2} as it is implemented in the neutrino event generator GENIE \cite{GENIE} by integrating over $E_{\pi}$ up to 3.0 GeV and averaged over the MINERvA flux. The results are shown as dotted curves in Fig. \ref{fig:ds_dcos}. In both cases, CC and NC, our results prefer small angles. We note that the integrated Rein-Seghal cross section in Fig. \ref{fig:ds_dcos} for CC is 61$\%$ larger than our cross section and has a long tail; for the NC case the Rein-Seghal is 48$\%$ larger. As we mentioned earlier, the SciBooNE experiment gave only an upper bound for CC coherent scattering; in addition to that it also reported event distributions for various angles. For $\theta< 35^{\circ}$ there is an excess of events above the estimated background concentrated at small values of $\phi$ (see Fig. 6 in  \cite{Hiraide}).  In addition to that our results are consistent within two standard deviations with the bounds reported by SciBooNE experiment \cite{ScibooneNC, Sciboone}.
  
On the experimental frontier the MINERvA experiment, with a fully active detector, is searching for coherent scattering \cite{Palomino,Higuera},  and is expected that it will bridge the gap between the recent measurements at low energy and early measurements at higher energies on a wide range of nuclear targets. Liquid argon TPCs (Time Projection Chambers) should be able to observe nuclear stubs and eliminate most incoherent backgrounds from coherent pion production searches, as was done in bubble chamber experiments. At the end we can expect multiple experiments searching for coherent scattering with multiple detection techniques and many nuclei as targets. They will  obtain precise measurements of coherent scattering, and will test the models we have discussed.

\section*{Acknowledgments}
We wish to thank Dr. J. G. Morfin for helpful discussion. A. Higuera gratefully acknowledges the support of graduate  grant award by CONACyT M\'exico and Universidad de Guanajuato. 

\section*{Appendix}
Using the equation given in section $IV$, it is possible to compute the Jacobian 
 \begin{equation}\label{Jacobian}
J(\theta, \theta', \phi) = det \left |\begin{array}{ccc}
\frac{\partial Q^{2}}{\partial \theta'} & \frac{\partial Q^{2}}{\partial \theta} & \frac{\partial Q^{2}}{d\phi} \\
\frac{\partial t}{\partial \theta'} & \frac{\partial t}{\partial \theta} & \frac{\partial t}{d\phi} \\
\frac{\partial \alpha}{\partial \theta'} & \frac{\partial \alpha}{\partial \theta} & \frac{\partial \alpha}{d\phi}
\end{array}
\right| \notag
\\ 
\end{equation}


\begin{align}
 J(\theta, \theta', \phi)  = \frac{4EE^{2}_{\pi}}{E-E_{\pi}}|\vec{q}| |\vec{p}_{\pi}| \frac{sin\theta'}{cos\alpha} \bigg \{  
 \bigg(-sin\theta cos\theta'  + cos\theta sin\theta' cos\phi \bigg) 
 \left( \frac{sin\theta cos\phi }{sin\zeta} - \frac{sin\theta sin\phi}{sin^{2}\zeta} cos\zeta \frac{\partial \zeta}{\partial \phi} \right) \\ \notag
 +  sin\theta sin\theta' sin\phi \left( \frac{cos\theta sin\phi}{sin\zeta} - \frac{sin\theta sin\phi}{sin^{2}\zeta} cos\zeta \frac{\partial \zeta}{\partial \theta}\right)\bigg \},
 \end{align}

 with
\begin{equation}
 \frac{\partial \zeta}{\partial \theta'} = \frac{cos\theta sin\theta' - sin\theta cos\theta' cos\phi}{\sin\zeta},
 \end{equation}
 \begin{equation}
\frac{\partial \zeta}{\partial \theta} =  \frac{sin\theta cos\theta' - cos\theta sin\theta' cos\phi}{\sin\zeta},
\end{equation}
\begin{equation}
\frac{\partial \zeta}{\partial \phi}  =  \frac{sin\theta sin\theta' sin\phi}{\sin\zeta}. 
 \end{equation}
 In order to appreciate the contribution of the various terms to the cross section we computed the elements of the density matrix. The results in Table \ref{table} indicate that $L_{l0}$ and $L_{ll}$ are indeed smaller and in addition they are multiplied by factors that are of order 1 or smaller.
 \begin{center}
 \captionof{table}{Numerical calculation for density-matrix elements, with $E$ = 3.0 and $E_{\pi}$ 0.4 GeV.} \label{table} 
 \begin{tabular}{cccc}
 \hline
$\ Q^{2} \  (GeV/c)^{2} \ $ &  $\ \ L_{00\ \ }$  &  $\ \ L_{l0}\ \ $  &   $\ \ L_{ll}\ \ $  \\
\hline
0.005 &	2.620 &	0.517&	0.144 \\
0.010 &	6.164 &	0.558&	0.094 \\
0.015 &	9.534 &	0.565&	0.077 \\
0.030 &	18.602 &	0.558&	0.061\\
0.050 &	28.677 &	0.5369&	0.054\\
0.070&     36.987 &	0.515&	0.051\\
0.100 &	47.035&	0.486&	0.049\\
0.130 &	54.979&	0.461&	0.048\\
0.150 &	59.408&	0.446 &	0.047\\
0.180 &	65.057&	0.426&	0.046\\
\hline
 \end{tabular}
 \end{center}

\end{document}